\documentclass[5p,twocolumn]{elsarticle}
\usepackage{graphicx,latexsym}
\usepackage{dcolumn}
\usepackage{amssymb,amsmath,bm}
\usepackage{subfigure}
\usepackage{braket}
\usepackage{siunitx}
\usepackage{array}
\usepackage{float}
\usepackage[nopar]{lipsum}
\usepackage{caption}
\usepackage{multirow}
\usepackage{bigstrut}
\usepackage[numbers]{natbib}

\usepackage[utf8]{inputenc}
\usepackage[english]{babel}
\usepackage{amsmath}
\usepackage{amsthm}

\usepackage{epstopdf}

\biboptions{sort&compress}

\usepackage[super]{nth}

\newcommand{\angstrom}{\text{\normalfont\AA}}
\usepackage{hyperref}
\hypersetup{
    pdfnewwindow=true,       
    colorlinks=true,         
    linkcolor=blue,          
    citecolor=blue,          
    filecolor=magenta,       
    urlcolor=black           
}

\usepackage[normalem]{ulem}

\def\sec#1{Sec.\ \ref{#1}}

\def\fig#1{Fig.\ \ref{#1}}

\journal{}

\begin{document}

\begin{frontmatter}


\title{First-Principles Study of Structural, Electronic, Thermal, and Optical Properties of Quasi-2D C$_2$N$_2$O Using GGA and HSE06}

\author[a1]{Hemn Gharib Hussein}
\ead{hemn.hussein@univsul.edu.iq}
\address[a1]{Physics Department, College of Education, University of Sulaimani, Sulaimani 46001, Kurdistan Region, Iraq}

\author[a2]{Nzar Rauf Abdullah}
\address[a2]{Division of Computational Nanoscience, Physics Department, College of Science,
	\\ University of Sulaimani, Sulaimani 46001, Kurdistan Region, Iraq}

\author[a3]{Vidar Gudmundsson}
\address[a3]{Science Institute, University of Iceland, Dunhaga 3, IS-107 Reykjavik, Iceland}


\begin{abstract}

DFT and AIMD are used to investigate the structural, stability, electronic, thermal, and optical properties of the quasi-2D C$_2$N$_2$O structure. The structure exhibits thermal and energy stability, signifying robustness under ambient conditions, however less dynamical stability is observed.
The electronic structure investigation reveals that C$_2$N$_2$O displays semiconducting properties with a moderate indirect band gap resulting from the hybridisation of $p$-orbitals of N, C, and O atoms, with band gap values of 2.3 eV (GGA) and 3.9 eV (HSE06). The optical properties, including the dielectric function, optical conductivity, and refractive index, are thoroughly analyzed to clarify the electronic transitions. The material exhibits considerable optical absorption in the visible and ultraviolet spectrum, with notable anisotropy between in-plane and out-of-plane polarizations. Furthermore, plasmon resonance occurs at around 3.8 eV, relating to the collective oscillations of charge carriers. The thermal properties indicate a heat capacity of around 382 J/mol$.$K at 300 K, which is close to and slightly above the standard Dulong-Petit limit for this structure, indicating near-complete excitation of lattice vibrational modes at room temperature. The lattice thermal conductivity is extremely low, reaching approximately 0.017 W/m.K at 300 K, primarily attributed to significant phonon scattering, evidenced by a scattering rate of roughly 3.2 $1/ps$ in the phonon frequency ranges. The findings demonstrate that the C$_2$N$_2$O structure maintains structural stability while allowing for tunable electronic, optical, and thermal properties, making it a promising candidate for nanoscale optoelectronic and thermal control applications.

\end{abstract}


\end{frontmatter}

\section{Introduction}

Two-dimensional (2D) carbon-based compounds have garnered considerable scientific attention owing to their exceptional structural, electronic, optical, and thermal properties, facilitating a range of applications in thermoelectrics and nanoelectronics. Graphene, the first synthesised 2D material, exhibits remarkable thermal conductivity and carrier mobility; however, its application in semiconducting and optoelectronic devices is constrained by its zero band gap \cite {novoselov2004electric, lee2008measurement}. To address this limitation, many graphene derivatives, including carbon nitride (C$_2$N, C$_3$N$_4$, C$_3$N$_5$), and carbon oxynitride systems, have been suggested to impart semiconducting properties while maintaining chemical stability \cite {robertson2002diamond, mahmood2016two, zeng2019all, kim2018ordered, mane2017highly}. Carbon-nitride frameworks, such as g-C$_3$N$_4$, have emerged as attractive photocatalysts and electronic materials due to their adjustable band gaps, robust covalent C-N bonds, and advantageous thermal stability \cite {doi:10.1021/acsanm.3c04740, ghosh2021nitrogen, hussein2025buckling, fu2018g}.

Recently, carbon-nitrogen-oxygen (C-N-O) systems have garnered heightened interest due to the integration of oxygen into carbon-nitride frameworks, which effectively alters chemical bonding and electronic structures, facilitating adjustable stability and functional characteristics \cite {shen2023new, cui2015first}. There has been an increasing focus on hybrid carbon-nitride-oxide compounds due to their potential to combine the benefits of nitrogen and oxygen doping in carbon networks. The addition of oxygen atoms alters the electronic environment and local charge distribution, resulting in improved orbital hybridisation and the emergence of new phonon modes that customise both optical and thermal properties \cite {qiu2017one, huang2020oxygen}. Theoretical and experimental investigations have demonstrated that partial replacement of nitrogen with oxygen atoms in two-dimensional carbon-nitride frameworks effectively reduces the band gap, enhances dipole moment asymmetry, and improves visible-light absorption, offering opportunities for photocatalytic and thermoelectric applications \cite {das2024unveiling, roy2021graphitic}. Additionally, the presence of mixed C-N/O bonding introduces a degree of anharmonicity in the lattice, which, in turn, affects phonon dispersion and subsequently alters the heat transfer characteristics of these low-dimensional materials \cite {van2019thermal}.

This work presents a thorough first-principles examination of the structural, electronic, optical, and thermal characteristics of a unique quasi-2D carbon oxynitride structure, C$_2$N$_2$O, using density functional theory (DFT) in conjunction with ab initio molecular dynamics (AIMD) simulations. We examine the thermodynamic and dynamic stability of the structure, investigate its band structure and density of states, and assess its optical and phonon-related properties. Special emphasis is placed on its phonon band structure, heat capacity, lattice thermal conductivity, group velocity, and scattering rate, which together dictate the phonon-mediated heat transport characteristics. The findings elucidate the impact of coexisting C-N and C-O bonds, along with flat phonon bands, on heat transport. This study seeks to identify quasi-2D C$_2$N$_2$O as a viable option for nanoelectronic and thermoelectric applications, owing to its optimal blend of semiconducting properties, moderate thermal conductivity, and structural integrity.

This paper is organised as follows: Section \ref{Computational Details} details the computational technique, while \sec{Results} presents the geometry, stability, electronic, optical, and thermal characteristics of C$_2$N$_2$O structures. Section \ref {Conclusions} presents the conclusions.

\section{Computational Details}\label{Computational Details}

Using the Quantum ESPRESSO (QE) package, first-principles calculations based on DFT are done to look into the quasi-2D C$_2$N$_2$O structural, electronic, optical, and thermal properties \cite{giannozzi2009quantum, giannozzi2017advanced}. The exchange-correlation potential is treated within the Generalized Gradient Approximation (GGA) using the Perdew-Burke-Ernzerhof (PBE) functional \cite{perdew1996generalized}. To guarantee convergence, the plane-wave basis set is truncated at a kinetic energy cutoff of $ 816.3$ eV.  The Brillouin zone is sampled with a Monkhorst-Pack $7\times6\times6$ k-point grid for structure relaxation and self-consistent field ({\it scf}) calculations, while a more refined $21\times18\times18$ grid is employed for non-self-consistent field ({\it nscf}) computations to derive the electronic density of states (DOS). Computation results are graphically visualized through the use of the XCrySDen software \cite{kokalj2003computer}.
All atomic structures are completely optimised until the leftover atomic forces are less than $0.001$ eV$/\angstrom$ and with the total energy tolerance of $ 1.4 \times 10^{-4} $ eV. The band structure and optical characteristics are derived from the converged $scf$ charge density.

Phonon dispersion and thermodynamic parameters, including the phonon band structure and heat capacity, are computed by integrating QE with the PHONOPY package. Lattice thermal conductivity, phonon group velocities, and phonon scattering rate are calculated using the PHONO3PY code, which relies on third-order interatomic force constants and use a $q$-point mesh of $25\times25\times5$ \cite{togo2023implementation}.

\section{Results and Discussion}\label{Results}

In this section, we present geometric, stability, electronic, optical and thermal properties of quasi-2D C$_2$N$_2$O structure.

\subsection{Geometry}

\begin{figure}[hbt!]
	\centering
	\includegraphics[width=1\linewidth]{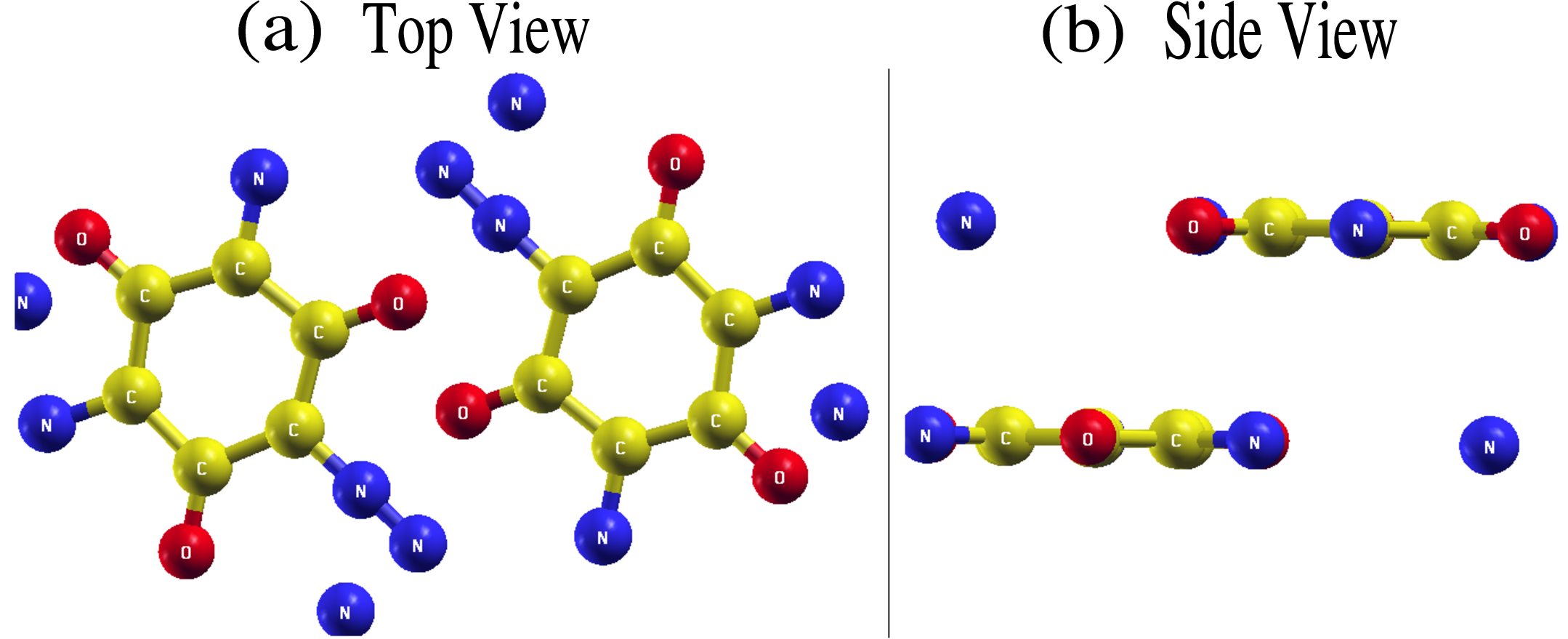}
	\caption{Crystal structure of quasi-2D C$_2$N$_2$O in (a) top view and (b) side view. Carbon, nitrogen, and oxygen atoms are colored yellow, blue, and red, respectively.}
	\label{figure01}
\end{figure}

This section presents the structural information of the quasi-2D C$_2$N$_2$O structure. The geometric configuration of C$_2$N$_2$O structure is demonstrated  in \fig{figure01}, where the atoms of carbon, nitrogen, and oxygen are shown as yellow, blue, and red, respectively. Figure \ref{figure01} presents both the top and side views of the structure, demonstrating that C$_2$N$_2$O is a two-dimensional (2D) layered structure, commonly known as a quasi-2D material with a primitive unit cell containing 30 atoms.  Additionally, it is found that the optimal bond lengths are 1.23 $\angstrom$ for C-O, 1.31 $\angstrom$ for C-N,  1.46 $\angstrom$ for C-C, and  1.13 $\angstrom$ for N-N. The optimised lattice parameters are a = 9.4407 $\angstrom$, b = 9.4407 $\angstrom$, and c = 6.1607 $\angstrom$, with lattice angles $\alpha = 90.0^{\circ}, \; \beta = 90.0^{\circ}, \; \text{and} \; \gamma = 120.0^{\circ}.$ These calculation results suggest a roughly hexagonal configuration with periodic stacking along the c-axis, consistent with the material's quasi-2D layered structure.

\subsection{Stability}

In this section, we use several methods to assess the stability of the quasi-2D C$_2$N$_2$O structure. To evaluate the energetic stability, the formation energy as a function of the lattice constant is calculated. Next, the phonon band structure is analyzed to verify dynamical stability. Finally, AIMD simulation is used to illustrate the thermal stability of the structure.

\begin{figure}[hbt!]

	\centering
	\includegraphics[width=1\linewidth]{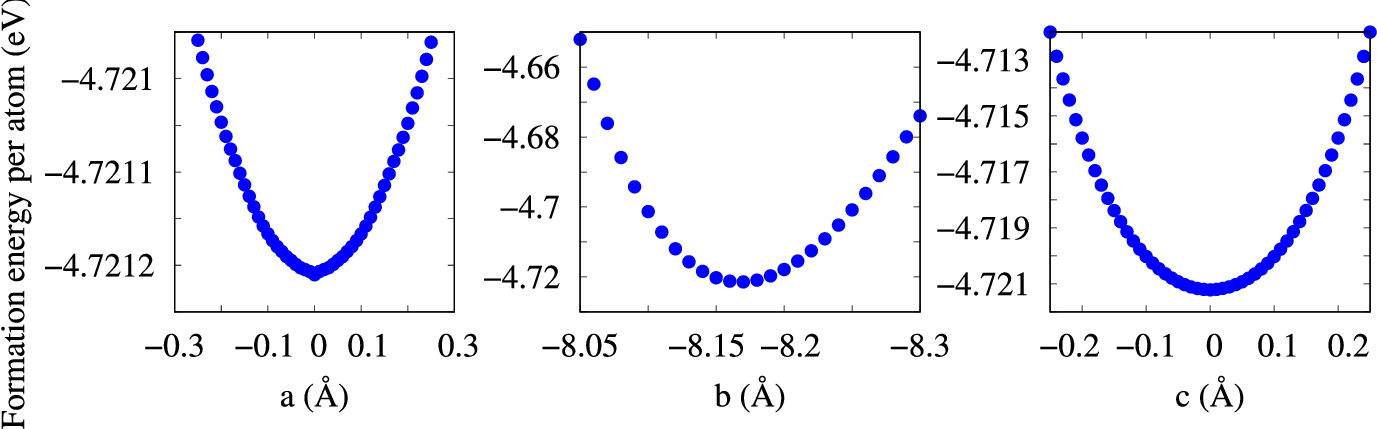}
	\caption{Formation energy of quasi-2D C$_2$N$_2$O as a function of lattice constant ($a$, $b$, and $c$).}
	\label{figure02}
\end{figure}

The formation energy serves as a crucial indicator of a material's energetic stability, defined as the disparity between the total energy of the optimized compound and the cumulative energies of its constituent atoms in their isolated states. A negative formation energy indicates that the material can spontaneously form and is energetically favorable, whereas a positive value suggests instability and a tendency to breakdown or react with other substances \cite {jehan2023insight}. The formation energy per atom (E$_f$) of quasi-2D C$_2$N$_2$O structure is calculated by using the expression,

\begin{equation*}
E_f = \frac{E_{\text{tot}}^{\mathrm{C}_2\mathrm{N}_2\mathrm{O}} - N_C \mu_{\mathrm{C}} - N_N \mu_{\mathrm{N}} - N_O \mu_{\mathrm{O}}}{N_{\text{total}}},
\end{equation*}

where $E_\mathrm{tot}^{C_2N_2O}$ is the total DFT energy of the quasi-2D C$_2$N$_2$O, $N_C$, $N_N$, and $N_O$ are the numbers of carbon, nitrogen, and oxgyen atoms typically from the stable elemental phases, respectively, $\mu_C$, $\mu_N$, and $\mu_O$ are the chemical potentials of carbon, nitrogen, and oxgyen, respectively, and N$_\mathrm{total}$ is the total number of atoms of the quasi-2D C$_2$N$_2$O structures. As illustrated in \fig{figure02}, the formation energy is plotted versus lattice constants $a$ (a), $b$ (b), and $c$ (c), and it reaches its minimal value at approximately -4.72 eV.
The calculated formation energy of the quasi-2D C$_2$N$_2$O structure is negative, indicating that it is energetically stable.
Under equilibrium conditions, C$_2$N$_2$O can exist as an energetically stable, as this minimum corresponds to the most stable configuration of the quasi-2D structure.

\begin{figure}[hbt!]
	\centering
	\includegraphics[width=0.7\linewidth]{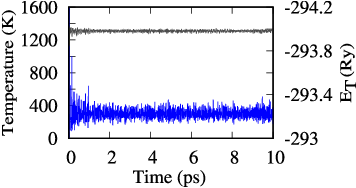}
	\caption{AIMD simulation of quasi-2D C$_2$N$_2$O structure.}
	\label{figure03}
\end{figure}

As shown in \fig{figure03}, AIMD simulations are carried at 300 K for a total duration of 10 ps using a time step of 1.0 fs in order to further evaluate the thermal stability of the C$_2$N$_2$O structure. The temperature profile shows very slight variations around the target value during the simulation, indicating that the system maintains thermal equilibrium in ambient settings \cite{abdullah2024novel}. Good energy conservation and dynamic stability are further supported by the practically constant total energy per atom (represented in gray color) and overall fluctuations in energy lees than 0.7 eV \cite{van2019thermal}. The simulation reveals no notable atomic rearrangements, bond breaks, or structural distortions, suggesting that the C$_2$N$_2$O framework maintains its stiffness and integrity in the face of heat disturbances. These results show that the quasi-2D C$_2$N$_2$O structure has good thermal stability at room temperature.

\begin{figure}[hbt!]
	\centering
	\includegraphics[width=0.6\linewidth]{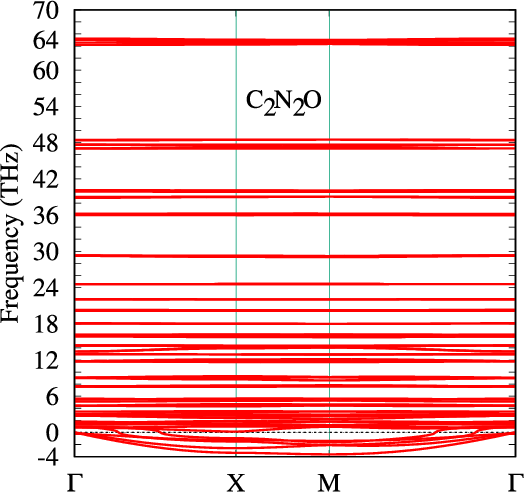}
	\caption{Phonon band structure of quasi-2D C$_2$N$_2$O.}
	\label{figure04}
\end{figure}

The phonon band structure of the quasi-2D C$_2$N$_2$O structure is determined by using the PHONOPY software inside the DFT framework in order to examine its dynamical stability. Figure \ref{figure04} demonstrates that the phonon band structure reveals some imaginary (negative) frequencies at the X and M points, signifying that the C$_2$N$_2$O structure is dynamically less stable. Furthermore, C$_2$N$_2$O exhibits flat phonon bands, indicating the presence of localized vibrational modes. The presence of flat bands signifies that phonons are not particularly efficient in heat transport within these modes. Moreover, flat bands may be linked to dynamic instability, especially when coupled with imaginary frequencies, as observed in C$_2$N$_2$O. Overall, the findings of the phonon dispersion show that the C$_2$N$_2$O structure suffers some dynamically instability and has flat phonon bands.

\subsection{ Electronic properties}

\begin{figure}[hbt!]
	\centering
	\includegraphics[width=1\linewidth]{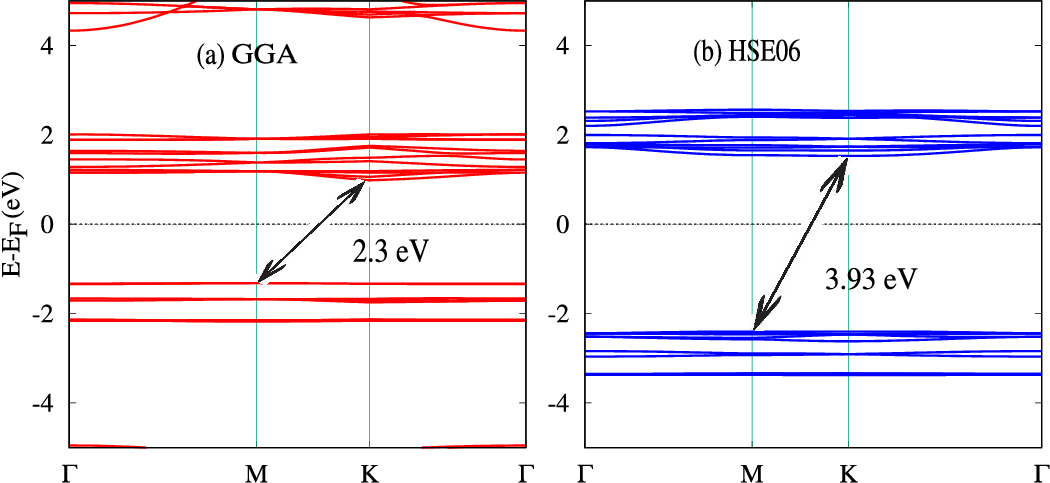}
	\caption{Electron band structure of quasi-2D C$_2$N$_2$O using GGA (a) and HSE06 (b) functional, where the Fermi energy is set to zero.}
	\label{figure05}
\end{figure}

\begin{figure}[hbt!]
	\centering
	\includegraphics[width=0.8\linewidth]{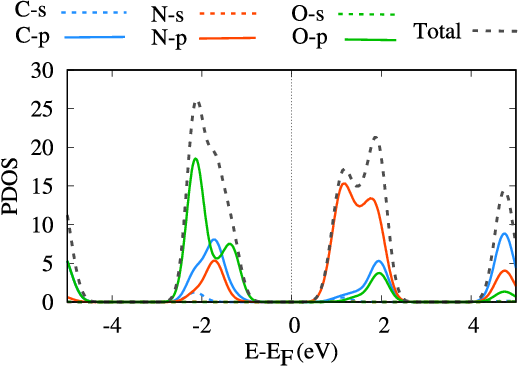}
	\caption{ Partial density of states of quasi-2D C$_2$N$_2$O.}
	\label{figure06}
\end{figure}

Figure \ref{figure05} and \fig{figure06} illustrate the electronic band structure and the partial density of states (PDOS) of the quasi-2D C$_2$N$_2$O structure, respectively. The Fermi energy ($E_f$) is set to zero. Under the GGA (PBE) approximation, C$_2$N$_2$O has an indirect band gap of around 2.3 eV, with the valence-band maximum (VBM) located at the M point and the conduction band minimum (CBM) located at the K point along the M-K path, as seen in \fig{figure05}(a). This confirms the material's indirect semiconducting nature.
For comparison, the band gap value is smaller than those reported for g-C$_3$N$_4$ ($\approx$ 2.7 eV) and p-C$_3$N$_4$ ($\approx$ 2.59 eV)  for GGA (PBE) approximation \cite{alaghmandfard2022comprehensive, huang2022band}. The HSE06 hybrid functional is also used for calculations to obtain a more accurate description of the electronic structure. This yields a large band gap of 3.93 eV for C$_2$N$_2$O as shown in \fig{figure05}(b). This outcome aligns with the established propensity of GGA-based functionals to underestimate band gap values. The margins of both the valence and conduction bands have a distinctly flat dispersion, indicating a substantial effective mass and hence restricted carrier mobility. The effective mass of electrons in the conduction band, as determined quantitatively,
is 0.445 $m_e$ using GGA, suggesting that the charge carriers are relatively light and flexible. The exceptionally flat valence-band edge around the VBM is reflected in the effective mass of holes in the valence band, which is -23.231 $m_e$ using GGA. The considerable magnitude indicates highly localized hole states and inferior hole mobility, while the negative sign represents the valence band's downward curvature.

Additional information on electronic characteristics is provided by the PDOS study, as seen in \fig{figure06}. The CBM comes primarily from the N-$p$ orbitals. In contrast, the VBM is dominated by the O-$p$ orbitals. This suggests that the main electronic transitions between O-$p$ and N-$p$ states take place across the band gap. The limited energy dispersion of these orbitals at the band borders confirms their localized nature, aligning with the observed flat bands in the band structure. Large effective masses, especially in the valence region, are the result of reduced band dispersion caused by these localized O-$p$ and N-$p$ states. All things considered, the PDOS and band structure analyses show that localized $p$-orbital interactions affect the electronic behaviour of C$_2$N$_2$O, giving it an indirect semiconducting structure with low hole mobility and comparatively higher electron transport. This makes it a promising candidate for upcoming optoelectronic applications.

\subsection{Optical properties}

The optical characteristics of the C$_2$N$_2$O structure are discussed in this section in order to have a better understanding of how it interacts with electromagnetic radiation. Together with the real part (Re($\varepsilon$)) and imaginary part (Im($\varepsilon$)), the complex dielectric function describes the material's optical response. Information on inter-band electronic transitions is provided by the Im($\varepsilon$) component, which corresponds to the absorption of electromagnetic energy, while the Re($\varepsilon$) component depicts the material's polarization and dispersion behavior under an external electric field. These characteristics are intimately linked to the material's electronic configuration and band structure. To ensure high precision, the dielectric functions were computed using the Independent Particle Approximation (IPA) \cite{ehrenreich1959self}.
This method provides a comprehensive image of the optical transitions in the C$_2$N$_2$O structure by capturing the interaction between light and both occupied and unoccupied.

\begin{figure}[hbt!]
	\centering
	\includegraphics[width=1\linewidth]{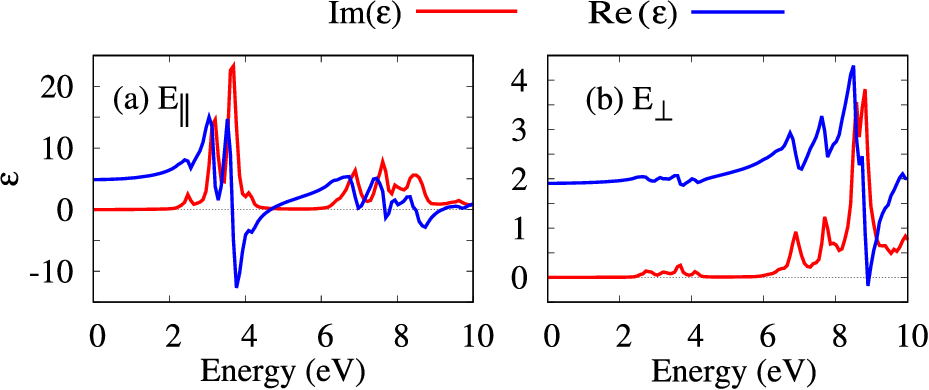}
	\caption{Im($\varepsilon$) (red) and Re($\varepsilon$) (blue) in the case of E$_{\parallel}$ (a) and E$_{\perp}$ (b).}
	\label{figure07}
\end{figure}

Figure \ref{figure07} illustrates the(Re($\varepsilon$)) and Im($\varepsilon$) components of the dielectric function for the electric field polarized parallel (E$_{\parallel}$) and perpendicular (E$_{\perp}$) to the C$_2$N$_2$O structure, respectively, throughout a photon energy spectrum of 0-10 eV. At low photon energies, the optical response is predominantly isotropic, especially in the imaginary component, Im($\varepsilon$). The static dielectric constants, Re($\varepsilon$(0)), are around 4.87 for in-plane polarization (E$_{\parallel}$) and 1.91 for out-of-plane polarization (E$_{\perp}$), indicating significantly greater in-plane polarizability and strong optical anisotropy characteristic of two-dimensional systems. Both Re($\varepsilon$($\omega$)) and Im($\varepsilon$($\omega$)) exhibit significant peaks in the visible range. The first major absorption peak in Im($\varepsilon$($\omega$)) appears approximately 2.3 eV with a maximum of around 3.2, arising from inter-band transitions between O-$p$ valence band and N-$p$ conduction band. This structure's flat bands ensure that the optical and electronic band gaps are nearly equal. These transitions, consistent with an indirect band gap of 2.3 eV, come from localized $p$-state excitation around the Fermi level, as verified by partial density of states analysis. Near 2.5 eV, the Re($\varepsilon$($\omega$)) curve shows high dispersion, suggesting a strong photon-electron interaction. The C$_2$N$_2$O structure exhibits significant optical activity across the visible range, underscoring its applicability for advanced optoelectronic applications, optical filters, and ultraviolet photo-detectors.

Moreover, other intriguing phenomenona connected to the imaginary and real parts of the dielectric function are plasmons. They are made up of a material's collectively oscillating valence or conduction electrons. The dielectric function's real component crosses zero and shifts from a positive to a negative sign at the plasmon resonance energy, while the imaginary part concurrently achieves its maximum. This behavior indicates the existence of plasmonic excitation \cite{raether2006excitation, egerton2011electron}. The distinctive plasmon response of the system is confirmed by the observation of a unique plasmon frequency in the case of the C$_2$N$_2$O structure at around 3.8 eV, where Im($\varepsilon$) reaches its maximal value as Real($\varepsilon$) sharply drops from positive to negative.

We also analyze the in-plane ($x$ and $y$) and out-of-plane ($z$) components of the dielectric function to evaluate the isotropy of the optical response. The in-plane components exhibit nearly identical behavior, indicating that the optical response of C$_2$N$_2$O is essentially isotropic within the layer. In contrast, the out-of-plane component deviates significantly from the in-plane trends (not shown)

\begin{figure}[hbt!]
	\centering
	\includegraphics[width=1\linewidth]{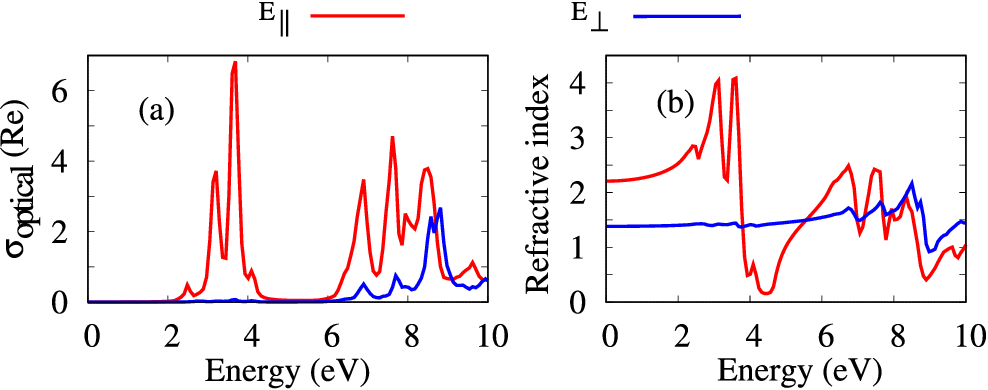}
	\caption{Optical conductivity (a) and refractive index (b) for E$_{\parallel}$ (red) and E$_{\perp}$ (blue) as a function of photon energy.}
	\label{figure08}
\end{figure}

Figure \ref{figure08} depicts the real part of the optical conductivity and the refractive index of the C$_2$N$_2$O structure for E$_{\parallel}$ and E$_{\perp}$ polarization across the photon energy spectrum from 0 to 10 eV. The optical conductivity, resulting from inter-band electronic excitation, stays essentially negligible below the band gap and significantly increases beyond 2.3 eV, aligning with the indirect electronic transition energy reported in the band structure. The initial significant conductivity peak appears between 3.0 and 3.5 eV, relating to photo-induced transitions between the O-$p$ and N-$p$ states that primarily govern the valence and conduction bands, respectively. The refractive index spectrum demonstrates significant dispersion in the visible range, with a constant value of roughly 2.1 at zero photon energy for E$_{\parallel}$. The refractive index diminishes progressively with rising energy, exhibiting typical dispersion characteristics, but retaining substantial values in the visible and ultraviolet (UV) spectrum, nearly between 3 and 4 eV, suggesting the viability of C$_2$N$_2$O for UV-visible optoelectronic and photonic applications. The concurrent rise in optical conductivity and refractive index in the visible spectrum highlights the material's effective light-matter interaction and its appropriateness for transparent conducting or light-modulating applications.

\subsection{Thermal properties}

This section delineates the thermal properties of the quasi-2D C$_2$N$_2$O structure, encompassing heat capacity, lattice thermal conductivity, phonon group velocity, and phonon scattering rate. The computations are executed utilising the PHONOPY package within the DFT framework, guaranteeing precise capture of the phonon contributions. Additionally, the PHONO3PY package is used to ascertain the lattice thermal conductivity and associated phonon transport parameters derived from third-order interatomic force constants \cite {togo2015first, togo2023implementation}.

\begin{figure}[hbt!]
	\centering
	\includegraphics[width=0.7\linewidth]{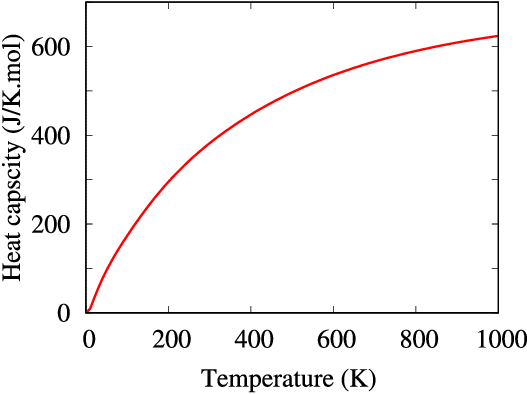}
	\caption{Heat capacity of quasi-2D  C$_2$N$_2$O structure as a function of temperature.}
	\label{figure09}
\end{figure}

Figure \ref{figure09} shows how the quasi-2D C$_2$N$_2$O structure's heat capacity changes with temperature between 0 and 1000 K. The material's ability to absorb and store thermal energy, which is mainly controlled by lattice vibrations, is reflected in its heat capacity. At low temperatures, the heat capacity escalates significantly with increasing temperature due to the gradual activation of low-energy phonon modes. This fast increase conforms to the Debye $C_v \propto T^3$ rule, typical of crystalline materials in the low-temperature domain. According to the \fig{figure09}, the total specific heat for C$_2$N$_2$O achieves a maximum value of 624 J/K.mol at 1000 K, while it reaches a value of 382 J/K.mol at about 300 K.
As the temperature rises, the slope of the curve gradually decreases and approaches saturation at high temperatures, signifying that the majority of phonon modes have been thermally excited. The physical correctness of the given data is validated by the fact that the specific heat per atom at 1000 K is 19.5 J/K.mol-atom, which is a bit far from the standard Dulong-Petit limit. The relatively lower heat capacity of quasi-2D C$_2$N$_2$O, compared with other 2D carbon-nitride materials 21.8 J/K.mol-atom for $\delta$-C$_3$N$_4$ and 21.1 J/K.mol-atom for carbon and boron nitrides at 1000 K, indicates a reduced capacity for thermal energy storage and dissipation \cite {fang2004lattice, tohei2015first}. This behaviour arises from its mixed C-N and C-O bonding network, which introduces several vibrational modes and mild anharmonicity, thereby enhancing phonon occupancy at increased temperatures. Therefore, the C$_2$N$_2$O structure demonstrates significant promise for temperature regulation and energy conversion in nanoelectronic and thermoelectric devices.

\begin{figure}[hbt!]
	\centering
	\includegraphics[width=0.65\linewidth]{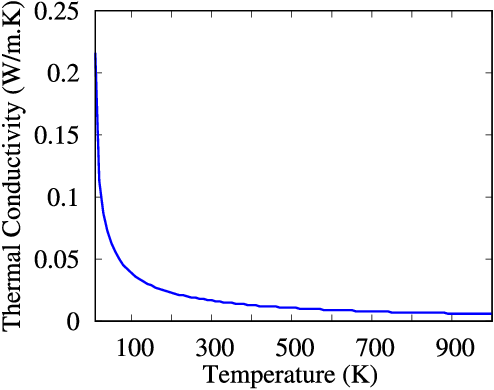}
	\caption{Lattice thermal conductivity for quasi-2D C$_2$N$_2$O as a function of temprtature.}
	\label{figure10}
\end{figure}

Thermal conductivity is another important thermal property, representing a material’s ability to transfer heat through phonon propagation within its crystal lattice \cite {plata2017efficient}. Figure \ref{figure10} illustrates the relationship between temperature and lattice thermal conductivity. The thermal conductivity of quasi-2D C$_2$N$_2$O decreases dramatically with increasing temperature, which is typical behavior for most crystalline solids. This reduction arises from enhanced phonon-phonon scattering at high temperatures, where intensified an-harmonic interactions among vibrational modes hinder the efficient transfer of thermal energy. Interestingly, the phonon band structure shown in \fig{figure04} reveals the presence of a flat phonon band, indicating a large effective mass that generally reduces lattice thermal conductivity. Consequently, despite its somewhat unstable dynamics, C$_2$N$_2$O has very low thermal conductivity due to the effects of optical phonon modes and the hybridization of C-N and C-O bonds, which facilitate partial phonon transit inside the lattice and phonon-phonon interactions. These findings suggest the structure is a promising candidate for thermal insulation applications in nanoscale electronic devices.

\begin{figure}[hbt!]
	\centering
	\includegraphics[width=0.8\linewidth]{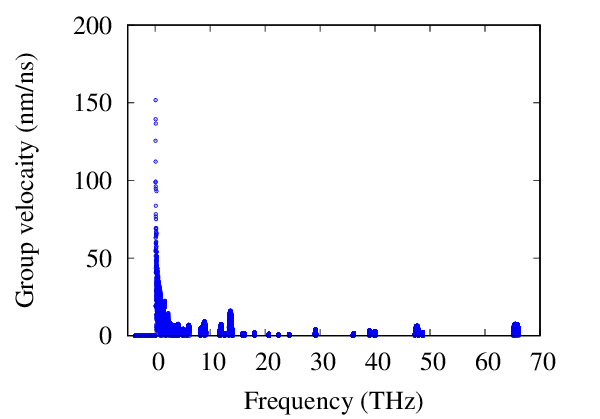}
	\caption{Phonon group velocity as a function of frequency for quasi-2D C$_2$N$_2$O.}
	\label{figure11}
\end{figure}

The phonon group velocity is a crucial determinant of heat transport in crystalline materials which can further confirm
the validity of lattice thermal conductivity.
It specifies the speed at which vibrational energy disseminates across the lattice as phonon wave packets. It can also be understood as the material's effective sound speed, which establishes a direct connection between macroscopic heat transfer and atomic vibrations \cite {luo2020vibrational, togo2023first}. The collective contribution of all phonon modes determines the lattice thermal conductivity ($\kappa$), which can also be written as $\kappa = \sum_n C_n v_{g,n}^2 \tau_n$, where $C_n$, $v_{g,n}$, and $\tau_n$ stand for the heat capacity, group velocity, and lifetime of the $n$-th phonon mode, respectively.
Figure \ref{figure11} illustrate phonon group velocity of the quasi-2D C$_2$N$_2$O as a function of frequency. The findings indicate that the group velocity reaches its peak at low frequencies and progressively diminishes with rising phonon frequency, a trait typical of two-dimensional materials. Low-frequency acoustic phonons predominantly govern heat transport, while high-frequency optical modes contribute less significantly due to their reduced group velocities and shorter lifetimes. The comparatively flat phonon band seen in \fig{figure04} signifies substantial phonon effective masses and diminished vibrational dispersion, resulting in decreased group velocities in some regions of the Brillouin zone. The simultaneous presence of flat and relatively dispersive branches facilitates partial energy transmission via acoustic modes, preserving a moderate total heat conduction capacity. The C$_2$N$_2$O structure shows a critical frequency dependency of phonon transport, with a maximum group velocity of around 152 nm/ns at zero frequency and a fall to about 8 nm/ns at 66 THz. This pattern demonstrates that low-energy acoustic phonons dominate thermal transport in C$_2$N$_2$O, but high-frequency phonon propagation is constrained by the flat optical branches' heavier vibrational nature. These characteristics work together to produce a balanced phonon transport mechanism, which supports the moderate lattice thermal conductivity for C$_2$N$_2$O that has been described.

\begin{figure}[hbt!]
	\centering
	\includegraphics[width=0.75\linewidth]{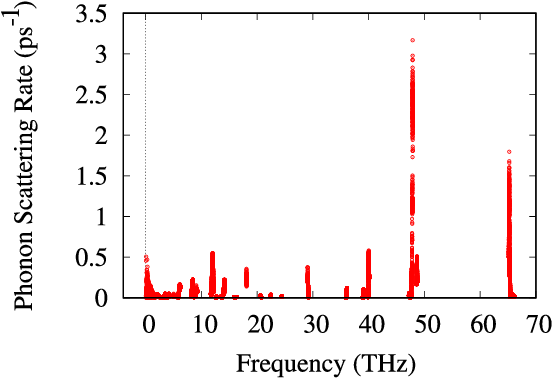}
	\caption{Phonon scattering rate of quasi-2D C$_2$N$_2$O as a function of frequency.}
	\label{figure12}
\end{figure}

The phonon scattering rate is another factor that affects the lattice thermal conductivity, which measures how frequently phonons lose momentum during lattice vibrations. When $\Gamma_n$ is the phonon scattering rate, it is inversely proportional to the phonon lifetime $\tau_n = 1/\Gamma_n$. The relation between phonon scattering rate and phonon frequency for the quasi-2D C$_2$N$_2$O structure is shown in \fig {figure12}. The findings indicate that the scattering rate rises with increasing phonon frequency, signifying intensified phonon–phonon interactions in the high-frequency domain. This behavior principally results from increased anharmonicity and the complex connection between acoustic and optical phonon branches within the mixed C-N-O bonding network. The scattering rate is comparatively low in the low-frequency range, facilitating extended phonon lifetimes and enhanced thermal conduction via acoustic modes.
The presence of flat phonon bands in \fig {figure04} enhances scattering effects by limiting phonon dispersion and promoting localized vibrational states, hence reducing coherence across phonon modes. Therefore, the augmented phonon scattering at higher frequencies substantially influences the temperature-dependent reduction in lattice thermal conductivity, as seen in \fig {figure10}. These findings demonstrate that low-frequency acoustic phonon propagation and high-frequency phonon scattering compete to control thermal transport in C$_2$N$_2$O, producing low stable thermal conductivity.

\section{Conclusions}\label{Conclusions}

In summary, first-principles calculations using DFT have been performed to investigate the structural, electronic, thermal, and optical properties of the quasi-2D C$_2$N$_2$O structure. The optimized geometry verifies that the structure is stable both energetically and thermally. The electronic band structure indicates a semiconducting characteristic with an indirect band gap mostly resulting from the hybridisation of $p$-orbitals of C, N, and O atoms. The thermal analysis indicates that lattice thermal conductivity diminishes with rising temperature owing to enhanced phonon–phonon scattering, while heat capacity converges to a constant value at elevated temperatures, in accordance with the Dulong-Petit limit. The optical examination demonstrates considerable asymmetry between parallel and perpendicular photon polarization, showing strong absorption in the ultraviolet range and little activity in the visible range. The significant plasmon resonance at around 3.8 eV is attributed to collective oscillations of charge carriers, confirming its applicability in optoelectronic applications. The structural durability, suitable band gap, and beneficial optical response suggest that the quasi-2D C$_2$N$_2$O structure is a promising candidate for future applications in nanoelectronics, and photonics devices.

\section{Acknowledgment}

This project is financed by the University of Sulaimani.  The calculations were partially conducted using resources from the Computational Nanoscience Lab/Research and Development Center at the University of Sulaimani.
The present work forms part of a PhD study conducted at the University of Sulaimani.

\section*{CRediT authorship contribution statement}

{\bf Hemn Gharib Hussein}:  Writing – original draft, Visualization, Validation, Software, Methodology, Investigation, Formal analysis, Data curation, Conceptualization.
{\bf  Nzar Rauf Abdullah}: Writing, review and editing -- original draft, Validation, Investigation, Formal analysis, Software,  Conceptualization, Supervision.
{\bf  Vidar Gudmundsson}: Writing, review and editing -- original draft, Validation, Investigation, Formal analysis, Conceptualization, Supervision.


\begin{thebibliography}{10}
	\expandafter\ifx\csname url\endcsname\relax
	\def\url#1{\texttt{#1}}\fi
	\expandafter\ifx\csname urlprefix\endcsname\relax\def\urlprefix{URL }\fi
	\expandafter\ifx\csname href\endcsname\relax
	\def\href#1#2{#2} \def\path#1{#1}\fi



\bibitem{novoselov2004electric}
K.~S. Novoselov, A.~K. Geim, S.~V. Morozov, D.-eng Jiang, Y.~Zhang,
S.~V. Dubonos, I.~V. Grigorieva, A.~A. Firsov,
Electric Field Effect in Atomically Thin Carbon Films,
Science 306~(5696) (2004) 666--669.

\bibitem{lee2008measurement}
C.~Lee, X.~Wei, J.~W. Kysar, J.~Hone,
Measurement of the Elastic Properties and Intrinsic Strength of Monolayer Graphene,
Science 321~(5887) (2008) 385--388.

\bibitem{robertson2002diamond}
J.~Robertson,
Diamond-like amorphous carbon,
Materials Science and Engineering: R: Reports 37~(4--6) (2002) 129--281.

\bibitem{mahmood2016two}
J.~Mahmood, E.~K. Lee, M.~Jung, D.~Shin, H.-J. Choi, J.-M. Seo, S.-M. Jung,
D.~Kim, F.~Li, M.~S. Lah, et al., Two-dimensional polyaniline (C$_3$N) from carbonized organic single crystals in solid state, Proceedings of the National Academy of Sciences 113~(27) (2016) 7414--7419.

\bibitem{zeng2019all}
J.~Zeng, Z.~Chen, X.~Zhao, W.~Yu, S.~Wu, J.~Lu, K.~P. Loh, J.~Wu,
From All-Triazine C$_3$N$_3$ Framework to Nitrogen-Doped Carbon Nanotubes:
Efficient and Durable Trifunctional Electrocatalysts,
ACS Applied Nano Materials 2~(12) (2019) 7969--7977.

\bibitem{kim2018ordered}
I.~Y. Kim, S.~Kim, X.~Jin, S.~Premkumar, G.~Chandra, N.-S. Lee,
G.~P. Mane, S.-J. Hwang, S.~Umapathy, A.~Vinu,
Ordered Mesoporous C$_3$N$_5$ with a Combined Triazole and Triazine Framework
and Its Graphene Hybrids for the Oxygen Reduction Reaction (ORR),
Angewandte Chemie 130~(52) (2018) 17381--17386.

\bibitem{mane2017highly}
G.~P. Mane, S.~N. Talapaneni, K.~S. Lakhi, H.~Ilbeygi, U.~Ravon,
K.~Al-Bahily, T.~Mori, D.-H. Park, A.~Vinu,
Highly Ordered Nitrogen-Rich Mesoporous Carbon Nitrides and Their Superior Performance
for Sensing and Photocatalytic Hydrogen Generation,
Angewandte Chemie International Edition 56~(29) (2017) 8481--8485.

\bibitem{doi:10.1021/acsanm.3c04740}
S.~Chandrappa, S.~J. Galbao, A.~Furube, D.~H. K. Murthy,
Extending the Optical Absorption Limit of Graphitic Carbon Nitride Photocatalysts: A Review,
ACS Applied Nano Materials 6~(21) (2023) 19551--19572.

\bibitem{ghosh2021nitrogen}
A.~Ghosh, H.~Saini, A.~Sarkar, P.~Guha, A.~K. Samantara, R.~Thapa,
S.~Mandal, A.~Mandal, J.~N. Behera, S.~K. Ray, et al.,
Nitrogen vacancy and hydrogen substitution mediated tunable optoelectronic properties
of g-C$_3$N$_4$ 2D layered structures: Applications towards blue LED to broad-band photodetection,
Applied Surface Science 556 (2021) 149773.

\bibitem{hussein2025buckling}
H.~G. Hussein, N.~R. Abdullah, V.~Gudmundsson,
Buckling strain-enhanced thermal insulation in C$_3$N$_4$ monolayers:
DFT and AIMD studies of electronic, optical and thermal properties,
Functional Materials Letters (2025) 2551060.

\bibitem{fu2018g}
J. Fu, J. Yu, C. Jiang, and B. Cheng,
g-C$_3$N$_4$-Based Heterostructured Photocatalysts,
Advanced Energy Materials,
Wiley Online Library, 2018.


\bibitem{shen2023new}
J.~Shen, Q.~Duan, J.~Miao, S.~He, K.~He, W.~Dai, C.~Lu,
New carbon--nitrogen--oxygen compounds as high energy density materials,
Chinese Physics B 32~(9) (2023) 096302.

\bibitem{cui2015first}
J.~Cui, S.~Liang, X.~Wang, J.~Zhang,
First principle modeling of oxygen-doped monolayer graphitic carbon nitride,
Materials Chemistry and Physics 161 (2015) 194--200.


\bibitem{qiu2017one}
P.~Qiu, C.~Xu, H.~Chen, F.~Jiang, X.~Wang, R.~Lu, X.~Zhang,
One step synthesis of oxygen doped porous graphitic carbon nitride with remarkable
improvement of photo-oxidation activity: Role of oxygen on visible light photocatalytic activity,
Applied Catalysis B: Environmental 206 (2017) 319--327.

\bibitem{huang2020oxygen}
J.~Huang, H.~Wang, H.~Yu, Q.~Zhang, Y.~Cao, F.~Peng,
Oxygen Doping in Graphitic Carbon Nitride for Enhanced Photocatalytic Hydrogen Evolution,
ChemSusChem 13~(18) (2020) 5041--5049.

\bibitem{das2024unveiling}
S.~K. Das, L.~Patra, P.~Samal, P.~K. Sahoo,
Unveiling the Reactivity of Oxygen and Ozone on C$_2$N Monolayer,
Physica Status Solidi (RRL) -- Rapid Research Letters 18~(12) (2024) 2400148.

\bibitem{roy2021graphitic}
P.~Roy, A.~Pramanik, P.~Sarkar,
Graphitic Carbon Nitride Sheet Supported Single-Atom Metal-Free Photocatalyst
for Oxygen Reduction Reaction: A First-Principles Analysis,
The Journal of Physical Chemistry Letters 12~(11) (2021) 2788--2795.

\bibitem{van2019thermal}
H.~van Gog, W.-F. Li, C.~Fang, R.~S. Koster, M.~Dijkstra, M.~van Huis,
Thermal stability and electronic and magnetic properties of atomically thin
2D transition metal oxides,
npj 2D Materials and Applications 3~(1) (2019) 18.

\bibitem{perdew1996generalized}
J.~P. Perdew, K.~Burke, M.~Ernzerhof,
Generalized Gradient Approximation Made Simple,
Physical Review Letters 77~(18) (1996) 3865.

\bibitem{giannozzi2009quantum}
P.~Giannozzi, S.~Baroni, N.~Bonini, M.~Calandra, R.~Car, C.~Cavazzoni,
D.~Ceresoli, G.~L. Chiarotti, M.~Cococcioni, I.~Dabo, et al.,
QUANTUM ESPRESSO: a modular and open-source software project for quantum simulations of materials,
Journal of Physics: Condensed Matter 21~(39) (2009) 395502.

\bibitem{giannozzi2017advanced}
P.~Giannozzi, O.~Andreussi, T.~Brumme, O.~Bunau, M.~Buongiorno Nardelli,
M.~Calandra, R.~Car, C.~Cavazzoni, D.~Ceresoli, M.~Cococcioni, et al.,
Advanced capabilities for materials modelling with Quantum ESPRESSO,
Journal of Physics: Condensed Matter 29~(46) (2017) 465901.

\bibitem{kokalj2003computer}
A.~Kokalj,
Computer graphics and graphical user interfaces as tools in simulations of matter at the atomic scale,
Computational Materials Science 28~(2) (2003) 155--168.

\bibitem{togo2023implementation}
A.~Togo, L.~Chaput, T.~Tadano, I.~Tanaka,
Implementation strategies in phonopy and phono3py,
Journal of Physics: Condensed Matter 35~(35) (2023) 353001.

\bibitem{jehan2023insight}
A.~Jehan, M.~Husain, S.~Bibi, N.~Rahman, V.~Tirth, A.~Azzouz-Rached,
M.~Y. Khan, M.~Nasir, K.~Inayat, A.~Khan, et al.,
Insight into the structural, optoelectronic, and elastic properties of
AuXF$_3$ (X = Ca, Sr) fluoroperovskites: DFT study,
Optical and Quantum Electronics 55~(14) (2023) 1242.

\bibitem{abdullah2024novel}
N.~R. Abdullah, B.~J. Abdullah, H.~G. Hussein, V.~Gudmundsson,
Novel ZnO nanosheet with buckling stress: First principles study of electronic,
structural stability, phonon vibrations, lattice thermal and optical conductivity,
Chemical Physics Letters 844 (2024) 141269.

\bibitem{van2019thermal}
H.~van Gog, W.-F. Li, C.~Fang, R.~S. Koster, M.~Dijkstra, M.~van Huis,
Thermal stability and electronic and magnetic properties of atomically thin
2D transition metal oxides,
npj 2D Materials and Applications 3~(1) (2019) 18.

\bibitem{ehrenreich1959self}
H.~Ehrenreich, M.~H. Cohen,
Self-Consistent Field Approach to the Many-Electron Problem,
Physical Review 115~(4) (1959) 786.

\bibitem{raether2006excitation}
H.~Raether,
Excitation of Plasmons and Interband Transitions by Electrons,
Springer, 2006.

\bibitem{egerton2011electron}
R.~F. Egerton,
Electron Energy-Loss Spectroscopy in the Electron Microscope,
Springer Science \& Business Media, 2011.

\bibitem{togo2015first}
A.~Togo, I.~Tanaka,
First principles phonon calculations in materials science,
Scripta Materialia 108 (2015) 1--5.


\bibitem{fang2004lattice}
C.~M. Fang and G.~A. De Wijs,
Lattice vibrations and thermal properties of carbon nitride with defect ZnS structure from first-principles calculations,
Journal of Physics: Condensed Matter,
IOP Publishing, 2004.

\bibitem{tohei2015first}
T. Tohei, H.-S. Lee, and Y. Ikuhara,
First Principles Calculation of Thermal Expansion of Carbon and Boron Nitrides Based on Quasi-Harmonic Approximation,
Materials Transactions,
The Japan Institute of Metals and Materials, 2015.

\bibitem{plata2017efficient}
J.~J. Plata, P.~Nath, D.~Usanmaz, J.~Carrete, C.~Toher, M.~de Jong,
M.~Asta, M.~Fornari, M.~Buongiorno Nardelli, S.~Curtarolo,
An efficient and accurate framework for calculating lattice thermal conductivity
of solids: AFLOW–AAPL Automatic Anharmonic Phonon Library,
npj Computational Materials 3~(1) (2017) 45.

\bibitem{luo2020vibrational}
Y.~Luo, X.~Yang, T.~Feng, J.~Wang, X.~Ruan,
Vibrational hierarchy leads to dual-phonon transport in low thermal conductivity crystals,
Nature Communications 11~(1) (2020) 2554.

\bibitem{togo2023first}
A.~Togo,
First-Principles Phonon Calculations with Phonopy and Phono3py,
Journal of the Physical Society of Japan 92~(1) (2023) 012001.

\bibitem{alaghmandfard2022comprehensive}
A.~Alaghmandfard, K.~Ghandi,
A Comprehensive Review of Graphitic Carbon Nitride (g-C$_3$N$_4$)-Metal Oxide-Based
Nanocomposites: Potential for Photocatalysis and Sensing,
Nanomaterials 12~(2) (2022) 294.

\bibitem{huang2022band}
M.~Huang, Z.~Ai, L.~Xu, K.~Zhang, Z.~Kong, Y.~Shao, Y.~Wu, X.~Hao,
Band structure-controlled P-C$_3$N$_4$ for photocatalytic water splitting via
appropriately decreasing oxidation capacity,
Journal of Alloys and Compounds 895 (2022) 162513.


\end{thebibliography}
\end{document}